\documentstyle[12pt]{article}
\begin{document}

\hfill \vbox{\hbox{INLO-PUB-02/01}
             \hbox{FSU TPI-04/01}}
\begin{center}{\Large\bf Classical Gauge Vacua as Knots}\\[2cm] 

{\bf Pierre van Baal}\\
{\em Instituut-Lorentz for Theoretical Physics, University of Leiden,\\
P.O.Box 9506, NL-2300 RA Leiden, The Netherlands}\\[6mm]

{\bf Andreas Wipf} \\
{\em Friedrich-Schiller-Universit\"at Jena, Theoretisch-Physikalisches
Institut,\\ Max-Wien-Platz 1, D-07743 Jena, Germany}\\[10mm]

\end{center}

\section*{Abstract}

The four dimensional $O(3)$ non-linear sigma model introduced by Faddeev and 
Niemi, with a Skyrme-like higher order term to stabilise static knot solutions
classified by the Hopf invariant, can be rewritten in terms of the complex 
two-component $CP_1$ variables. A further rewriting of these variables in 
terms of $SU(2)$ curvature free gauge fields is performed. This leads us to 
interpret $SU(2)$ pure gauge vacuum configurations, in a particular maximal 
abelian gauge, in terms of knots with the Hopf invariant equal to the winding 
number of the gauge configuration.

\section{Introduction}

In this Letter we address some simple results that involve rewriting the
Faddeev-Niemi model~\cite{Faddeev:1997zj}. This model has stable static 
solutions that represent knots. Much work has been invested in interpreting 
this model as an effective low-energy representation of $SU(2)$ gauge 
theory~\cite{Faddeev:1999eq,Langmann:1999nn,Cho:1999wp} and investigating 
the quality of this approximation by inverse Monte Carlo techniques~\cite{Dit}.
This interpretation in part is motivated by 't~Hooft's notion of abelian 
projection~\cite{'tHooft:1981ht}.

The model is defined in terms of a space-time dependent vector $\vec n(x)$ 
of fixed (here chosen unit) length. To allow for non-trivial static solutions
a Skyrme-like higher-order term is added~\cite{Faddeev:1975tz}, through the 
introduction of a composite gauge field strength $F_{\mu\nu}(x)=\frac{1}{2}
\vec n(x)\cdot(\partial_\mu\vec n(x)\wedge\partial_\nu\vec n(x))$. Note that 
with $\vec n(x)$ a unit three-vector, $\partial_\mu\vec n(x)$ is perpendicular 
to $\vec n(x)$, and $\partial_\mu\vec n(x)\wedge\partial_\nu\vec n(x)$ is 
proportional to $\vec n(x)$. The factor of proportionality is precisely 
$2F_{\mu\nu}(x)$. Thus, one also has $F^2_{\mu\nu}(x)=\frac{1}{4}(\partial_\mu
\vec n(x)\wedge\partial_\nu\vec n(x))^2$. The action is given by
\begin{equation}
S=\int d^4x\left(m^2\partial_\mu\vec n(x)\cdot\partial^\mu\vec n(x)
-\frac{1}{2e^2}F_{\mu\nu}(x)F^{\mu\nu}(x)\right).\label{eq:O3}
\end{equation}
By rescaling $x$ with $(em)^{-1}$, $e^2S$ becomes independent of both $e$ and 
$m$. With this understood, we will now put $e=m=1$. Finite energy requires 
$\vec n(\vec x)$ to approach a constant vector at spatial infinity. In this
way static configurations are classified by the topological maps from $S^3$ 
into $S^2$, characterised by the Hopf invariant. The two-form $F(\vec x)=
\vec n(\vec x)\cdot(d\vec n(\vec x)\wedge d\vec n(\vec x))$ implicitly 
defines an abelian gauge field one-form $A(\vec x)$ through $F(\vec x)=
dA(\vec x)$, in terms of which the Hopf invariant is given by 
$Q=\frac{1}{4\pi^2}\int A(\vec x)\wedge F(\vec x)$. Remarkably, the 
energy is bounded by a fractional power of this Hopf 
invariant~\cite{Vakulenko:1979uw,Kundu:1982bc}.
\begin{equation}
E=\int d^3x\left((\partial_i\vec n(x))^2+\frac{1}{2}F_{ij}^2(x)\right)
\geq c|Q|^{3/4},
\end{equation}
with $c=16\pi^2 3^{3/8}\sim238$. This gives a rough bound, which can be 
improved on~\cite{Ward:1998pj} (by roughly a factor 2). Extensive numerical 
studies~\cite{Battye:1998pe,Hietarinta:2000ci} have gone up 
to $Q=8$, with energies indeed following the fractional power of $Q$.

\section{The $CP_1$ formulation}

We first discuss the reformulation in terms of $CP_1$ fields, as well-known 
from two dimensions~\cite{Eichenherr:1978qa,D'Adda:1978uc}. The main advantage 
is that the abelian gauge field involved in defining the Hopf invariant, no 
longer needs to be defined implicitly. To be specific, one introduces a 
complex two-component field $\Psi(x)$. The two degrees of freedom associated 
to the $n$ field are obtained by identifying any two $\Psi$'s which differ by 
an overall nonvanishing complex scale factor. This is achieved by constraining 
$\Psi$ to have unit length, and introducing local abelian gauge invariance, 
obvious from the following relation to the $n$ field:
\begin{equation}
n^a(x)=\Psi^\dagger(x)\tau^a\Psi(x),
\end{equation}
where $\tau^a$ are the Pauli-matrices. The abelian gauge invariance of the 
$CP_1$ model leads to a composite gauge field
\begin{equation}
A_\mu(x)=-i\Psi^\dagger(x)\partial_\mu\Psi(x),
\end{equation}
and one verifies by direct computation that indeed $F(x)=dA(x)$. 
Useful identities for these computations are the completeness relation
$\delta_{ij}\delta_{kl}+\tau^a_{ij}\tau^a_{kl}=2\delta_{il}\delta_{jk}$
and $i\varepsilon_{abc}\tau^b_{ij}\tau^c_{kl}=\tau^a_{kj}\delta_{il}-
\tau^a_{il}\delta_{jk}$. For the action, Eq.~(\ref{eq:O3}), we find the 
following result
\begin{equation}
S=\int d^4x\left(4(D_\mu\Psi)^\dagger(x)D^\mu\Psi(x)-\frac{1}{2}
F_{\mu\nu}(x)F^{\mu\nu}(x)\right),\label{eq:CP1}
\end{equation}
where $D_\mu=\partial_\mu-iA_\mu(x)$ is the covariant derivative. Note that 
$\Psi^\dagger(x)D_\mu\Psi(x)=0$ and that the energy density can be written 
as a square, $E=\int d^3x~|(2D_i+B_i(\vec x))\Psi(\vec x)|^2$.

\section{The $SU(2)/U(1)$ formulation}

The next reformulation makes use of the fact that any two-component complex 
vector of unit length is in one to one relation to an $SU(2)$ group element.
Alternatively we can write $\Psi(x)=g(x)\Psi_0$. For convenience we choose 
$\Psi^\dagger_0=(1,0)$, such that
\begin{equation}
n_a(x)=\frac{1}{2}{\rm tr}\left(\tau_3 g^\dagger(x)\tau_a g(x)\right). 
\end{equation}
As we will see, the winding number of $g(\vec x)$ as a map from $R^3$ to 
$SU(2)$ is precisely the Hopf invariant.  This observation is in itself 
not new~\cite{Battye:1998pe}. But we will push it a little further here.

We introduce currents $J_\mu^a(x)$ through $J_\mu(x)=i\tau_a J_\mu^a(x)=
g^\dagger(x)\partial_\mu g(x)$. A simple calculation shows that
\begin{equation}
A_\mu(x)=J_\mu^3(x)\quad\mbox{and}\quad\partial_\mu\Psi^\dagger(x)\partial^\mu
\Psi(x)=J_\mu^a(x)J_a^\mu(x).
\end{equation}
We can interpret the currents just as well as components of an $SU(2)$ gauge 
connection, which is pure gauge, $G(x)=dJ(x)+J(x)\wedge J(x)=0$, with 
$J(x)\equiv J_\mu(x)dx_\mu$. For later use we also introduce 
$J^a(x)\equiv J^a_\mu(x)dx_\mu$. In particular in components, we have
$G^3_{\mu\nu}(x)=\partial_\mu J^3_\nu(x)-\partial_\nu J^3_\mu(x)-2
(J_\mu^1(x)J^2_\nu(x)-J_\nu^1(x)J^2_\mu(x))=0$. It leads to the useful
identity
\begin{equation}
F(x)=dJ^3(x)=2J^1(x)\wedge J^2(x)\quad\mbox{or}\quad
F_{\mu\nu}(x)=2(J_\mu^1(x)J^2_\nu(x)-J_\nu^1(x)J^2_\mu(x)).
\end{equation}
With the help of this relation it is now also easy to show that the Hopf 
invariant is exactly equal to the winding number of the gauge function 
$g(\vec x)$, 
\begin{equation}
\frac{1}{4\pi^2}A(\vec x)\wedge F(\vec x)=
\frac{1}{2\pi^2}J^3(\vec x)\wedge J^1(\vec x)\wedge J^2(\vec x)=
\frac{1}{24\pi^2}{\rm tr}(g^\dagger(\vec x)
dg(\vec x))^3,
\end{equation}
which can of course also be related to the non-abelian Chern-Simons form,
\begin{equation}
\frac{1}{4\pi^2}A(\vec x)\wedge F(\vec x)=-\frac{1}{8\pi^2}{\rm tr}\left(
J(\vec x)\wedge dJ(\vec x)+\frac{2}{3}J(\vec x)\wedge J(\vec x)\wedge 
J(\vec x)\right).
\end{equation}
A similar relation between the Hopf invariant and a non-Abelian Chern-Simons
form was discussed in Ref.~\cite{Langmann:1999nn}. But we wish to argue here
that the static solitons of the Faddeev-Niemi model actually represent 
classical Yang-Mills vacua in a non-linear maximally abelian gauge. It is
this result that we believe to be new.

First we note that,
\begin{equation}
(D_\mu\Psi)^\dagger(x)D^\mu\Psi(x)=\partial_\mu\Psi^\dagger(x)\partial^\mu
\Psi(x)-A_\mu(x)A^\mu(x)=J_\mu^1(x)J_1^\mu(x)+J_\mu^2(x)J_2^\mu(x),
\end{equation}
which makes the $SU(2)/U(1)$ nature of the action explicit, since both terms
in Eq.~(\ref{eq:CP1}) can be written in terms of just $J^1_\mu(x)$ and 
$J^2_\mu(x)$.  So the energy of a static configuration is given in terms of 
the "charged" components of the non-abelian gauge field only
\begin{equation}
E=\int d^3x\Biggl(4\left(J^1_i(\vec x)J^1_i(\vec x)+J^2_i(\vec x)J^2_i(\vec x)
\right)+2\left(J^1_i(\vec x)J^2_j(\vec x)-J^1_j(\vec x)J^2_i(\vec x)\right)^2
\Biggr)
\end{equation}
The first term agrees exactly with the functional that defines the maximal 
abelian gauge, by minimising along the gauge orbit, leaving the abelian 
subgroup generated by $\tau_3$ unfixed~\cite{'tHooft:1981ht,Kronfeld:1987ri}. 
This remains true for the full energy functional, which can thus just as well 
be interpreted as the gauge fixing functional for a non-linear maximal abelian 
gauge. As the three parametrisations are mathematically equivalent, we are 
entitled to interpret the minima of the energy functional in the sector with 
a given value of $Q$ as gauge fixed pure gauge (i.e. curvature free, or flat) 
connections in a sector with gauge field winding number $Q$. Therefore, there 
is a gauge fixing in terms of which the gauge vacua with different winding 
number can be characterised by inequivalent knots.

\section{Conclusions}

In the light of the attempts to relate the Faddeev-Niemi model to full 
non-abelian gauge theory, our result is a rather sobering one, even 
though it also involves an abelian projection. Within the context of 
our interpretation, there seems not much need to address the quantum 
fluctuations. It should, however, be noted that at the quantum level 
the three models are not equivalent, as the path integral measure 
depends on the chosen representation. It is the measure that seems 
to cause some of the problems in relating the Faddeev-Niemi model to 
the full $SU(2)$ gauge theory.

We hope this Letter provides inspiration for new ways of viewing the 
topological non-trivial nature of non-abelian gauge theories. The relation 
of pure gauge theory vacua to knots is also suggestive from the point of 
view of Chern-Simons theory and topological field theory. Instantons become 
knot changing operations, as also suggested in Ref.~\cite{Langmann:1999nn}, 
and one may even hope the present results can have some mathematical 
ramifications~\cite{Witten:1989hf,vanBaal:1990aw}. We will leave this 
to future studies.

\section*{Acknowledgements}

P.v.B thanks Leo Stodolsky, Valya Zakharov and Dieter Maison for discussions 
and hospitality at the MPI in Munich, where this work was first presented. 
A.W. thanks the Lorentz-Institute in Leiden, where much of the work has been 
done, for excellent working conditions and Falk Bruckmann for many useful 
discussions.

\end{document}